\documentclass[reprint,amsmath,amssymb,superscriptaddress,nofootinbib,aps]{revtex4-2}

\usepackage{xcolor}
\usepackage{graphicx}
\usepackage{braket}
\usepackage{caption}
\usepackage{subcaption}
\usepackage{hyperref}
\usepackage{xr}
\usepackage{amsthm}
\usepackage{mathtools}

\newtheorem{Def}{Definition}
\newtheorem{Prop}{Proposition}
\newtheorem{The}{Theorem}

\theoremstyle{definition}

\begin{document}

\title{A simple algorithm to reflect through eigenspaces of unitaries}

\author{Baptiste Claudon}\email{baptiste.claudon@qubit-pharmaceuticals.com}
\affiliation{Qubit Pharmaceuticals, Advanced Research Department, 75014 Paris, France}
\affiliation{Sorbonne Universit\'e, LJLL, UMR 7198 CNRS, 75005 Paris, France}
\affiliation{Sorbonne Universit\'e, LCT, UMR 7616 CNRS, 75005 Paris, France}
\date{\today}

\maketitle

\section*{Abstract}

Reflections are omnipresent tools in quantum algorithms. We consider the task of reflecting through the eigenspace of an implementable unitary. Such reflections are generally designed using phase estimation or linear combination of unitaries. These methods have size and depth that scale favorably with the desired precision and the spectral gap of the unitary. However, they require a number of ancilla qubits that grows with both parameters. Here, we present a simple algorithm with the same size and depth scaling but requiring only a single ancilla qubit for all problem instances. As such, this algorithm is expected to become the reference method to reflect through eigenspaces of unitaries. 

\section{Introduction}

Due to the exponential speedup of quantum algorithms over the best known classical methods in solving certain problems, quantum computing recently gained a lot of attention \cite{365700, Low2019hamiltonian}. Numerous quantum algorithms rely on reflections through well-chosen subspaces. Quantum amplitude estimation \cite{Brassard_2002}, amplitude amplification \cite{Yoder_2014}, search problems \cite{grover1996fastquantummechanicalalgorithm} and Markov chain algorithms \cite{1366222} have complexities that are even quantified in terms of the number of uses of these reflection operators. Such reflections were initially constructed using controlled state preparation oracles \cite{grover2002creatingsuperpositionscorrespondefficiently, Chiang2009EfficientCF} and controlled phase gates \cite{polylog, khattar2024riseconditionallycleanancillae}, or the phase estimation algorithm \cite{Kitaev:1995qy}.\\

In the present work, we consider the task of constructing a reflection through the eigenspace of an implementable unitary. To the best of our knowledge, the most efficient construction to reflect through one-dimensional eigenspaces is based on linear combination of unitaries \cite{Chowdhury2018ImprovedIO}. It provides the desired reflection up to a spectral norm error $\epsilon$ with circuit size and depth complexity $\mathcal O\left(\delta^{-1}\log\left(1/\epsilon\right)\right)$, where $\delta$ is an angular gap separating the target eigenvalue from the others. The construction requires a number of ancilla qubit that increases with both $1/\epsilon$ and $1/\delta$.\\

Here, we present a construction based on the quantum signal processing framework \cite{PRXQuantum.5.020368, sunderhauf2023generalized}. Our construction also presents circuit size and depth complexities $\mathcal O\left(\delta^{-1}\log\left(1/\epsilon\right)\right)$ but only requires a single ancilla qubit, independently of $\epsilon$ and $\delta$. Our result is summarized in Proposition \ref{prop:main_result}.\\

Because it only uses a single ancilla qubit, our decomposition is expected to become the routine of choice to construct reflections through eigenspaces of unitaries. It could ultimately improve the efficiency of quantum Monte-Carlo methods \cite{akhalwaya2023modularenginequantummonte}, of interest in financial modeling \cite{glasserman2004monte}, or the preparation of physically relevant quantum states \cite{Lin2020}, with implications in quantum chemistry \cite{Whitfield10032011, Feniou_2024}. 

\section{Technical background}

The definitions and statements presented in this section are known. They originate from \cite{PRXQuantum.5.020368, sunderhauf2023generalized, 1366222}. We will implicitly consider operators acting on a quantum computational Hilbert space associated with a finite number of qubits. The $^\dag$ superscript denotes the complex conjugate and transposed operator. $\|\cdot\|$ denotes the spectral norm and $|\cdot|$ the modulus. $1$ denotes the identity operator. $B_r(0)$ denotes the centered open disc of radius $r>0$ in the complex plane and $\partial B_r(0)$ denotes its boundary. $\mathbb N=\{0, 1,...\}$ denotes the set of nonnegative integers. $\log$ denotes the logarithm in base $2$. $\ln$ denotes the natural logarithm in base $e$. $\sigma(U)$ denotes the spectrum of the operator $U$.

\subsubsection{Projected unitary encodings}

Let us start by introducing projected unitary encodings. They allow to implement normalized operators as the restriction of unitaries to certain input and output subspaces. The following definitions rely on the notion of partial isometries: operators whose restrictions to the space orthogonal to their kernels are isometries. Remark that if $\square$ is a partial isometry, then $\square\square ^\dag$ is an orthogonal projector and $\square^\dag\square=1$ on the orthogonal of its kernel.   

\begin{Def} Let $U$ be a unitary and $\square_L, \square_R$ be partial isometries. If $\square_L^\dag U\square_R=A$, for some operator $A$, then $\left(U, \square_L, \square_R\right)$ is said to be a Projected Unitary Encoding (PUE) of $A$.
\end{Def}

Note that if $\left(U, \square_L, \square_R\right)$ is a PUE of $A$, then necessarily $\|A\|\leq 1$.

\subsubsection{Generalized quantum signal processing}

Generalized Quantum Signal Processing (GQSP) is a framework to construct PUE of polynomials of unitaries. Every complex polynomial that maps all unitaries to operators of spectral norm less than $1$ can be implemented. The construction only requires access to controlled versions of the unitary and a single ancilla qubit. Theorem \ref{the:gqsp}, proven in \cite{PRXQuantum.5.020368}, will be sufficient for our purposes. 

\begin{The} Let $\Upsilon$ be a degree $d\in \mathbb N$ complex polynomial and $U$ be a unitary. If $\|\Upsilon(V)\|\leq 1$ for all unitaries $V$, then we can construct a PUE $(W(U), \ket0, 1)$ of 
\begin{equation}
\begin{pmatrix}
\Upsilon(U)\\
\Phi(U)
\end{pmatrix}
\end{equation} 
using $d$ controlled-$U$ operations, $\mathcal O(d)$ additional single-qubit gates and $1$ ancilla qubit. Here, $\Phi$ is any degree $d$ complex polynomial such that for all unitaries $V$:
\begin{equation}
\Upsilon(V)\Upsilon(V)^\dag+\Phi(V)\Phi(V)^\dag=1.
\label{eq:defining_complementary}
\end{equation}
\label{the:gqsp}
\end{The}

Note that the existence of $\Phi$ is a consequence of the Fejér-Riesz theorem \cite{alma991019855769703276, berntson2024complementarypolynomialsquantumsignal}.

\section{Results}

\subsubsection{Problem statement}

Let $U$ be a unitary acting on a finite set of qubits. Let $e^{i\theta}\in \sigma(U)$ be an eigenvalue of $U$ and $\Pi$ be the orthogonal projection on the corresponding eigenspace. We assume that we know a $\delta\in ]0, \pi[$ such that $\{\lambda\in ]\theta-\delta, \theta+\delta[:e^{i\lambda}\in \sigma(U)\}=\{\theta\}$. We are interested in constructing a PUE of the reflection $2\Pi-1$, up to spectral norm error $\epsilon\in]0, 1[$. For that purpose, we allow ourselves the use of ancilla qubits, arbitrary two-qubit gates, controlled-$U$ and controlled-$U^\dag$ operations. Up to considering $e^{-i\theta}U$ instead of $U$, we will assume without loss of generality that $\theta=0$. 

\subsubsection{Strategy}

Let $\Upsilon$ be a degree $d\in \mathbb N$ polynomial with complex coefficients and such that $\|\Upsilon(V)\|\leq1$ for every unitary operator $V$. Let $\Phi$ be any polynomial satisfying Equation \ref{eq:defining_complementary}. Note that $-\Phi$ also satisfies Equation \ref{eq:defining_complementary}. Theorem \ref{the:gqsp} therefore provides two PUEs $(W_{\pm}(U), \ket0, 1)$ of 
\begin{equation}
\begin{pmatrix}
\Upsilon(U)\\
\pm\Phi(U)
\end{pmatrix}.
\end{equation}
Then, notice that $\left(W_-(U)^\dag W_+(U), \ket0, \ket0\right)$ is a PUE of $2\Upsilon(U)\Upsilon(U)^\dag-1$. Indeed,
\begin{equation}
\begin{split}
\bra0\left(W_-(U)^\dag W_+(U)\ket0\right)&=\begin{pmatrix}
\Upsilon(U)^\dag & -\Phi(U)^\dag
\end{pmatrix}
\begin{pmatrix}
\Upsilon(U) \\ \Phi(U)
\end{pmatrix}\\
&=2\Upsilon(U)\Upsilon(U)^\dag-1,
\end{split}
\label{eq:pue}
\end{equation}
because Equation \ref{eq:defining_complementary} implies 
\begin{equation}-\Phi(U)\Phi(U)^\dag=\Upsilon(U)\Upsilon(U)^\dag-1.\end{equation}
We also used that polynomials of unitary operators are normal operators: $\Upsilon(U)\Upsilon(U)^\dag=\Upsilon(U)^\dag\Upsilon(U)$ and $\Phi(U)\Phi(U)^\dag=\Phi(U)^\dag\Phi(U)$. 

As a consequence, it is sufficient to find a polynomial $\Upsilon$ such that $\Upsilon(U)\approx\Pi$ to get an approximate PUE $\left(W_-(U)^\dag W_+(U), \ket0, \ket0\right)$ of the desired reflection $2\Pi-1$.

\subsubsection{Optimal complexity polynomial}

Fix $0<\epsilon<1$. Let us construct a polynomial $\Upsilon$ such that:
\begin{enumerate}
\item $\|\Upsilon(V)\|\leq1$ for all unitaries $V$, 
\item $\|\Upsilon(U)-\Pi\|\leq \epsilon$, and
\item $\Upsilon$ is of degree $\mathcal O\left(\delta^{-1}\log\left(1/\epsilon\right)\right)$.
\end{enumerate}
Let $t, n\geq1$ be integers and look for a polynomial of the form:
\begin{equation}
\Upsilon_{t, n}(x)=\left(\frac1t\sum_{k=0}^{t-1}x^k\right)^n.
\end{equation}
Clearly, $\Upsilon_{t, n}$ maps $B_1(0)$ to itself and can be implemented through the GQSP construction \ref{the:gqsp}. Also, $\Upsilon_{t, n}(1)=1$. We simply have to choose $t$ and $n$ so that for each for eigenvalue $e^{i\lambda}\neq1$ of $U$, $\left|\Upsilon_{t, n}\left(e^{i\lambda}\right)\right|\leq \epsilon$. By the geometric sum formula,
\begin{equation}
\Upsilon_{t, n}\left(e^{i\lambda}\right)^{1/n}=\frac 1t\frac{e^{i\lambda t}-1}{e^{i\lambda}-1}.
\end{equation}
Since $|e^{i\lambda t}-1|\leq2$ and $\left|e^{i\delta}-1\right|\leq\left|e^{i\lambda}-1\right|$:
\begin{equation}
\left|\Upsilon_{t, n}\left(e^{i\lambda}\right)\right|^{1/n}\leq \frac1t\frac{2}{\left|e^{i\lambda}-1\right|}\leq \frac1t\frac{2}{\left|e^{i\delta}-1\right|}.
\end{equation}
Letting $t=\left\lceil e/(2\left|e^{i\delta}-1\right|)\right\rceil$, we have that 
$
\left|\Upsilon_{t, n}\left(e^{i\lambda}\right)\right|^{1/n}\leq 1/e
$
for each eigenvalue $e^{i\lambda}\neq1$ of $U$. Letting $n=\lceil\ln\left(1/\epsilon\right)\rceil$ yields 
$
\left|\Upsilon_{t, n}\left(e^{i\lambda}\right)\right|\leq \epsilon
$
for every $e^{i\lambda}\in \sigma(U)\backslash\{1\}$.

\subsubsection{Final construction}

We now have a polynomial $\Upsilon$ such that $\|\Upsilon(U)-\Pi\|\leq \epsilon$. It follows that:
\begin{equation}
\begin{split}
&\left\|2\Upsilon(U)\Upsilon(U)^\dag-1-(2\Pi-1)\right\|=2\left\|\Upsilon(U)\Upsilon(U)^\dag-\Pi\Pi^\dag\right\|\\
&\leq 2\left\|\Upsilon(U)\|\|\Upsilon(U)^\dag-\Pi^\dag\right\|+2\left\|\Upsilon(U)-\Pi\|\|\Pi^\dag\right\|\\
&\leq 4\epsilon.
\end{split}
\end{equation}
Proposition \ref{prop:main_result} summarizes our finding.
\begin{Prop} The circuit $W_-(U)^\dag W_+(U)$ satisfies:
\begin{equation}
\left\|\bra0\left(W_-(U)^\dag W_+(U)\ket0\right)-(2\Pi-1)\right\|\leq 4\epsilon,
\end{equation}
and is constructed from
\begin{equation}
(t-1)n\in \mathcal O\left(\delta^{-1}\log\left(1/\epsilon\right)\right)
\end{equation}
controlled-$U$, controlled-$U^\dag$ and single-qubit gates.
\label{prop:main_result}
\end{Prop}

\section*{Acknowledgements}

The author would like to thank Pierre Monmarché, César Feniou, Jean-Philip Piquemal and Pablo Rodenas Ruiz for their careful reading of early versions of the manuscript and suggestions. 

\onecolumngrid
\bibliography{biblio}

\end{document}